\documentclass[aps,preprint,groupedaddress,showpacs]{revtex4}

\usepackage{graphicx}

\begin{document}

\title{Callen's Adiabatic Piston and the Limits of the Second Law of Thermodynamics}

\author{Bruno Crosignani}
\affiliation{Department of Applied Physics, California 
Institute of Technology, Pasadena,
California 91125, USA and \\ Dipartimento di Fisica, Universita' dell'Aquila, 67010 L'Aquila, Italy and \\
INFM-CNR, Universita' di Roma ``La Sapienza,'' 00185 Roma, Italy}

\author{Paolo Di Porto}
\affiliation{Dipartimento di Fisica, Universita' dell'Aquila, 67010 L'Aquila, Italy and \\
INFM-CNR, Universita' di Roma ``La Sapienza,'' 00185 Roma, Italy}

\author{Claudio Conti}
\affiliation{
Research center ``Enrico Fermi,'' Via Panisperna 89/A, Roma, Italy and \\
Research center SOFT INFM-CNR,\\ Universita' di Roma ``La Sapienza,''
00185 Roma, Italy
}

\date{\today}

\begin{abstract}
The  limits  of  the   Second  Law  of Thermodynamics,  which  reigns
undisputed  in   the  macroscopic  world,  are   investigated  at  the
mesoscopic  level, corresponding to spatial dimensions of a few microns.
An  extremely  simple isolated  system, modeled  after
Callen's adiabatic piston,  \cite{Callen60} can, under appropriate conditions, be described as a self-organizing
Brownian motor and shown to exhibit a perpetuum mobile behavior.
\end{abstract}

%%% PACS. 05.40.-a   05.70   05.70.L 05.40.J

\maketitle

\section{Introduction}
As first clearly  stated by Schroedinger ``a living  organism tends to
approach the  dangerous state of  maximum entropy, which is  death. It
can only keep  aloof from it, i.e., alive,  by continually drawing from
its  environment  negative  entropy.  What  an organism  fed  upon  is
negative entropy".\cite{Schroedinger45}  Is it possible  for an organism
immersed  in  a   thermal  bath  to  be  insulated,   and  thus  avoid
thermalization,   and   still  be   able   to   reduce  its   entropy?
Unfortunately,  the  above  scenario   contradicts  one  of  the  most
cherished laws of  physics, that is the Second  Law of Thermodynamics,
which has  victoriously resisted all  the attempts aimed at  finding a
particular  case where  it could  be  violated.  However,  one has  to
consider that  the Second  Law has been  formulated in the  context of
macroscopic  physics  and it  is  in this  context  that  it has  been
successfully  applied  and  verified.   While the  Second  Law,  being
inherently statistical in its nature, cannot carry over to microscopic
cases where the number of  involved particles is too small, its limits
of  validity  are not  well  understood  in  the transition region  bridging  the
macroscopic to the microscopic  world,
where the system dimensions are small but still the number of
particles
large enough to justify the use of Thermodynamics.

In particular,  some intermediate  mesoscopic regime, like,  e.g., the
biological  one associated  with  typical cell  dimensions (about  one
micron),  is still  \textit{terra incognita}  and open  to  possible surprises.
Recent  years  have  indeed   witnessed  a  growing  interest  in  the
Thermodynamics of  small-scale non-equilibrium devices,  especially in
connection  with   the  operation  and  the   efficiency  of  Brownian
motors.\cite{Reinmann02} These  are devices aimed  at extracting useful work  out of
thermal noise and are the  descendents of the famous ratchet mechanism
popularized  by Feynman \cite{FeynmanBook} in  order  to show  the impossibility  of
violating the Second Law  of Thermodynamics. From a quantitative point
of  view,  they  are  usually  described in terms of a variable  $x(t)$,
which  can be identified  as the  trajectory of  a ``Brownian
particle''  of   given  mass $M$ obeying  Newton's equation     of     motion.    
The     forces     acting    on     the ``particle'' are  those resulting from  some kind of
prescribed potential  $V(x)$, from the  viscous-friction term $-\gamma \dot{x}(t)$
and from  the randomly fluctuating  force (Langevin's force)
associated with  thermal noise.  Even in the  presence of  a spatially
asymmetric (ratchet-type) potential $V(x)$, no preferential direction of
motion is  possible and no perpetuum  mobile mechanism
leading to a violation of the second law.\cite{FeynmanBook}
 In this paper, we show that a very fundamental
thermodynamic     system,    the     so-called    ``adiabatic piston'',  
first introduced  by Callen,\cite{Callen60}  can  be actually  described as  a self-organizing Brownian
motor. Under specific conditions, the system can be analytically modeled as a harmonic oscillator undergoing Brownian motion.
%%%%%%%%%%%%%%%%%%%%%%%%%%%%%%%%%%%%%%%%
\begin{figure}
\centerline{\includegraphics[width=0.5\textwidth]{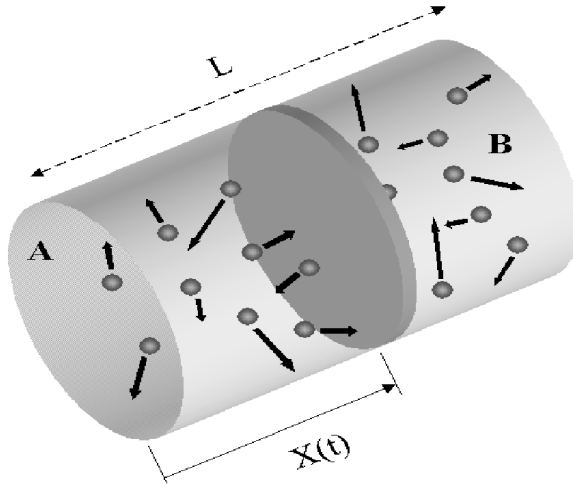}}
\caption{ The adiabatic piston: an
insulating cylinder divided into two regions by a movable,
frictionless and insulating piston.  \label{f1}}
\end{figure}
%%%%%%%%%%%%%%%%%%%%%%%%%%%%%%%%%%%%%%%%
The  relevant variable of  the system  is the  position $x(t)$  of the
moving wall with  respect to its central position  $L/2$, the harmonic
potential being not  prescribed but naturally emerging, as  well as the
friction term, as the result of the internal dynamics of the system.  

We recall  that   the  adiabatic-piston  problem  deals   with  a  system
consisting (see Fig.1)  of an adiabatic cylinder of  length $L$, divided
into  two sections  A  and B,  containing  the same  number  $n$ of  gas
moles  in different  equilibrium states,  by a  fixed frictionless
adiabatic piston .  At a given instant t=0, the piston  is let free to
move and the problem is to  predict the final equilibrium state of the
system.  This deceivingly  simple  question, which  cannot be  actually
answered in the frame  of equilibrium thermodynamics, has been firstly
solved in  1996 in  the frame of  a simple  kinetic model \cite{Crosignani96}  and has
since  then  given  rise  to   a  number  of  papers  of  increasing
sophistication. Both analytic \cite{Piasecki99,Chernov02,Munakata01,Gruber03} 
and molecular dynamics approaches \cite{Kestemont00,Renne01,White02,Mansour05} 
reveal the  existence of two
successive stages of evolution, the  first one leading the system to a
state of mechanical equilibrium  (where the pressures in both sections
are equal) and  the second, intrinsically random and  occurring over a
much  longer time  scale,  to a  final  state of  both mechanical  and
thermal equilibrium (where the average position of the piston is equal
to L/2  and the  average gas temperatures  in the two  sections become
equal).   The   above  conclusions   have  been  basically   drawn  by
investigating  the  average  behavior  of  the  significant  variables
describing the system evolution but less attention has been devoted to
the fluctuations  around these  average quantities, which  are somehow
assumed  to be small.  By extending  and clarifying  the results  of a
previous paper \cite{Crosignani01},  we wish to show  how this is
not  always the  case. More  precisely,  if one  starts from  specific
initial  conditions,  then  for  spatial dimensions  of  the  cylinder
corresponding to the mesoscopic range  (about 1 micron) and for values
of the ratio $\mu=M/M_g$ between the piston and the gas mass lesser than one,
fluctuations around equilibrium position can be a sizeable fraction of
the average  quantities, 
larger by a factor $\sqrt{M/m}$ ($m$ being the mass of the single gas molecule) than the ones expected if the system
were in thermal equilibrium.
This makes rather  questionable the statement
that in  this case a  final equilibrium configuration can  be actually
reached. Actually, the  evolution of the system from  an initial state
in  which both  $x(t)=X(t)-L/2$ and $\dot{x}(t)$ are zero  and the  temperatures are
equal  in the two  sections exhibits  unexpected features:  the system
fails to  settle down  to a well-defined  equilibrium state  since the
piston never comes to a  stop but keeps wandering symmetrically around
the initial  position,performing oscillations whose  mean-square value
(evaluated over an ensemble of  many replica) is much larger than that
pertaining  to  standard   thermal  fluctuations.  According  to  this
behavior, the  system appears  to deserve the  name of  {\it perpetuum
mobile},  even  if  there  is  no  preferential  direction  of  motion
($\langle  x\rangle  =\langle  \dot{x}\rangle  =0$).   
This unusual behavior can be associated with 
a possible challenge to the Second Law. In fact, a suitable process
can be conceived through which the entropy of the {\it isolated} 
adiabatic piston system turns out to undergo a significant decrease
(see Sections IV-V).
In practice, as we shall see below, no  significant  entropy   decrease  occurs 
outside  a mesoscopic  regime  corresponding  to  spatial  dimensions  of  a  few
microns;  in   fact,  the  time  interval  $t_{as}$   over  which  the
significant entropy decrease occurs depends on the fourth power of $L$
($t_{as}  \propto L^4$): thus,  any deviation  from the  above spatial
scale gives rise  to such large $t_{as}$ as to affect  the validity of our
model.  We  note that the coincidence  between the spatial  scales over
which  the  model is  valid  and  a  typical biological  dimension  is
definitely intriguing.
\section{The stochastic adiabatic piston}
The investigations  of the  adiabatic-piston dynamics appeared  in the
last ten years (see, e.g., \cite{Crosignani96,
Piasecki99,Chernov02,Munakata01,Gruber03}) 
have placed into evidence two
distinct  stages of  temporal evolution.  The first  one occurs,  on a
relatively short  time scale,  when the system  is let free  to evolve
from an initial configuration in which the two sections are in thermal
equilibrium at  different pressures, and  leads to a  final mechanical
equilibrium  (equal  pressures)  configuration  that  depends  on  the
initial conditions.  The second one  is characterized by  a stochastic
dynamic evolution  during which the  common pressure on the  two sides
remains  essentially constant,  and  may eventually  lead  to a  final
equilibrium  configuration in  which  also the  temperatures attain  a
common  value and  the  piston is  in  the central  position. In  this
section, we confine ourselves to the second stage by investigating the
piston dynamics starting from a specific initial configuration characterized
by $x(t=0)=\dot{x}(t=0)=0$ and by a common temperature $T_0$ in the two sections. In particular,
we  wish to  focus our  attention on  the fluctuations  of  the piston
positions around $L/2$. In effect, while $\langle x(t) \rangle$ and 
$\langle \dot{x}(t) \rangle$ turn out to
vanish at all times, as {\it a priori} dictated by symmetry, there is
a range of values of the system parameters, corresponding to
a mesoscopic regime where the mean-square value $\langle x^2(t)\rangle^{1/2}$ can be a
not  negligible  fraction  of  the  piston half-length  $L/2$,  a  quite
surprising  result which  seems to  indicate that  no  effective final
equilibrium  position can  be  reached.  

We  start  from the  equation
describing the deterministic piston motion in the presence of a finite
pressure difference, that is \cite{Crosignani96}
\begin{equation}
\label{eq1}
\ddot{X}=\displaystyle \frac{n R }{M}\left(\frac{T_A}{X}-\frac{T_B}{L-X} \right)-\sqrt{\frac{8 n R M_g}{\pi M^2}}
\left( \frac{\displaystyle\sqrt{T_A}}{X}+\frac{\displaystyle\sqrt{T_B}}{L-X}\right)\dot{X}+\frac{M_g}{M}
\left(\frac{1}{X}-\frac{1}{L-X}\right)\dot{X}^2
\end{equation}
where $R$,$M_g$ and $M$ are  the molar gas constant, the common values
of the gas  masses in $A$ and $B$, and  the piston mass, respectively,
while the dot stands for derivation with respect to time.  It has been
derived by assuming  that the velocity distributions of  the ideal gas
in the two sections are  Maxwellian, i.e., that thermal equilibrium is
continuously  restored   on  a  time  scale  much   shorter  than  the
characteristic time  scale of the  piston motion, an  assumption which
has to be  checked {\it a posteriori}.  We observe  that Eq.(1) is identical
to  the  one obtained  by  Munakata  and  Ogawa (Eq. (22)of
Ref. \cite{Munakata01})but  for the  extra-term  containing
$\dot{X}^2$. This  term,  which  is
negligible for finite pressure-difference  in the two sides, turns out
to play a fundamental role when studying the piston dynamics for equal
pressures,  i.e.,  in  the   situation  investigated  in  the  present
paper. In  this case, Eq.(\ref{eq1}) has  to be suitably  modified. In effect,
besides setting equal to zero the first term  on its R.H.S. associated with pressure
difference, we have to explicitly  take into account the random nature
of the hits suffered by the  piston because of the collisions with the
gas molecules. To this aim, we resort to Langevin's approach (see,
e.g. \cite{Crosignani01}) based on the introduction of an {\it ad hoc} Langevin acceleration
$a(t)$, so that Eq.(\ref{eq1}) is superseded by
\begin{equation}
\label{eq2}
\ddot{X}=-\frac{N}{M}\sqrt{\frac{8 k m }{\pi
}}\left(\frac{\displaystyle\sqrt{T_A}}{X}+
\frac{\displaystyle\sqrt{T_B}}{L-X} \right)\dot{X}+
\frac{ N m}{M}\left(\frac{1}{X}-\frac{1}{L-X}\right)\dot{X}^2 +a(t)\text{,}
\end{equation}
where $k$ is the Boltzmann constant and $N=M_g/m$ is the number of
molecules of mass $m$ on each side.

By taking again advantage of the equal pressure condition, i.e., $T_A(t)/X(t)=T_B(t)/[L-X(t)]=2T_0/L$ Eq.(\ref{eq2}) can be rewritten as
\begin{equation} \label{eq3}
\ddot{ X} + 
\frac{4N}{N}\sqrt{\frac{k T_0 m}{\pi  L}}
\left( \frac{1}{\sqrt{X}} + 
\frac{1}{\sqrt{L-X}} \right) \dot{X}
+ \frac{2 N m}{M} \left[ \frac{X-L/2}{X(L-X)} \right] \dot{X}^2= a(t)\text{.}
\end{equation}
Determining the  correct  expression for  $a(t)$ is  a
delicate task since the standard Langevin approach does not in general
carry over to nonlinear dynamical systems \cite{VanKampenBook}, as the one described by
Eq.(\ref{eq3}).  In  order  to be able to take  advantage of  the  Langevin  method,  we
linearize the  above equation:  a) by considering  small displacements
around the starting position of the piston, that is $|(X-L/2)|/(L/2)<<1$
and b)  by approximating  the square $\dot{X}^2$  of the  piston
velocity
by the average thermal value $k T_0/M$
; both hypotheses 
have to be proved consistent {\it a posteriori}. 
More precisely, by resorting to  assumption b) and by using Langevin's
method , it is possible  to identify two characteristic times, $t_{th}$ and
$t_{as}>>t_{th}$ (see  Eq.(\ref{eq10bis})). The first  one represents the  temporal scale
over which  the piston velocity  thermalizes, while the second  one is
the   characteristic  time  over   which  the   large  piston-position
fluctuations occur.  The relation $t_{th}<<t_{as}$ justifies  a posteriori the
application of the fluctuation-dissipation theorem.

After introducing the new variable $x(t)=X(t)-L/2$, the linearized version of Eq.(\ref{eq3}) reads
\begin{equation}\label{eq4}
\ddot{x}
+ \frac{4 N}{\sqrt{ \pi}}\frac{v_0}{L}\left(\frac{m}{M}\right) \dot{x}
                     + 4 N\frac{v_0^2}{L^2}\left(\frac{m}{M}\right)^2 x = a(t)\text{,}
\end{equation}
$v_0=(2k T_0/m)^{1/2}$ being the most probable velocity of the
corresponding gas Maxwellian distribution. \cite{SommerfeldBook}
At    this    point,   a    straightforward    application   of    the
fluctuation-dissipation theorem \cite{Pathria} yields
\begin{equation}
\langle a(t)a(t')\rangle =\frac{16 N}{\sqrt{\pi} L} v_0^3 \left(\frac{m}{M}\right)^2\delta(t-t')\text{.}  
\end{equation}
   
According to the above results, the variable $x(t)$,
describing the evolution of the state of our system, obeys an equation
formally similar to  that of a ``Brownian particle''
in one  dimension. However,  the potential term  and the  viscous drag
have naturally emerged out of  the internal dynamics of the system and
do not correspond to an external active force or to a phenomenological
friction term as in the case of Brownian motors. If we define
\begin{equation}
\label{eq5}
\beta=\frac{4 N}{\sqrt{\pi}}\frac{v_0}{L}\frac{m}{M}
\text{,}\hspace{1cm}\omega^2=4 N \frac{v_0^2}{L^2}\left(\frac{m}{M}\right)^2
\end{equation}
Eq.(\ref{eq4}) is formally identical to that describing the Brownian motion of
a harmonically-bound particle of mass $M$, that is
\begin{equation}
\label{eq6}
\ddot{x}+\beta\dot{x}+\omega^2 x=a(t)
\end{equation}
where  $a(t)$  is the  Langevin  acceleration,  a  case which  has  been
thoroughly investigated in the literature \cite{Chandrasekhar43}.

\section{The piston as a harmonically-bound Brownian particle}
The  solution  of  the  stochastic differential  equation  (\ref{eq6})
governing  the piston  motion  is  readily obtained  in  terms of  the
general results provided in \cite{Chandrasekhar43} for the Brownian motion of a
harmonically-bound particle. Equation (\ref{eq6}),together with
Eq.(\ref{eq5}),     describe    an     ``overdamped''     case    (in
fact, $\beta/\omega=\sqrt{16/\pi}\sqrt{N}>>1$, since $N$ is typically a
large number).For this situation, one has \cite{Chandrasekhar43}
\begin{equation}
\label{chandraeq}
\begin{array}{l}
\langle x(t)\rangle=x_0 e^{-\beta t /2}\left( \cosh \frac{\beta_1
t}{2}+\frac{\beta}{\beta_1} \sinh\frac{\beta_1 t}{2}
\right)+\frac{2u_0}{\beta_1}e^{-\beta t/2} \sinh \frac{\beta_1 t}{2}\text{,}\\
\langle u(t)\rangle=u_0 e^{-\beta t /2}\left( \cosh \frac{\beta_1
t}{2}-\frac{\beta}{\beta_1} \sinh\frac{\beta_1 t}{2}
\right)-\frac{2 x_0 \omega^2}{\beta_1}e^{-\beta t/2} \sinh \frac{\beta_1 t}{2}\text{,}\\
\langle x(t)^2 \rangle=\langle x(t) \rangle^2 +\frac{k
T_0}{M \omega^2}\left[1-e^{-\beta t}
\left(2\frac{\beta^2}{\beta_1^2}\sinh^2 \frac{\beta_1 t}{2} 
+\frac{\beta}{\beta_1} \sinh \beta_1 t +1\right)\right]\text{,}\\
\langle u(t)^2 \rangle=\langle u(t) \rangle^2 +\frac{k
T_0}{M }\left[1-e^{-\beta t}
\left(2\frac{\beta^2}{\beta_1^2}\sinh^2 \frac{\beta_1 t}{2} 
-\frac{\beta}{\beta_1} \sinh \beta_1 t +1\right)\right]\text{,}
\end{array}
\end{equation}
where $u(t)=dx/dt$ , $x_0=x(t=0)$, $u_0=u(t=0)$ and
\begin{equation}
\beta_1=\sqrt{\beta^2-4\omega^2}\cong \beta-2\omega^2/\beta\text{.}
\end{equation}
In our case $x_0=0$, $u_0=0$, so that, as expected, $\langle
x\rangle=\langle u \rangle=0$ at all times (see Eqs. (\ref{chandraeq})$_{1,2}$).
By taking the temporal asymptotic limit of the above equations, it is possible to recognize the existence of two characteristic time scales $t_{th}$ and $t_{as}$ , such that $t_{th}<<t_{as}$, given by
\begin{equation}
\label{eq10bis}
\begin{array}{l}
t_{th}=\frac{1}{2\beta}\text{,}\\
t_{as}=\frac{\beta}{2\omega^2}\text{.}
\end{array}
\end{equation}
They respectively represent the time over which the piston velocity thermalizes, i.e., its mean-square attains the value  
\begin{equation}
\label{eq10}
\langle \dot{x}^2 \rangle_{as}=\frac{k T_0}{M}\text{,} 
\end{equation}
and the time over which the mean-square value of the piston position reaches the asymptotic 
value
\begin{equation}
\langle x^2 \rangle_{as}=\frac{k T_0}{M \omega^2}\text{.}
\end{equation}
By using Eq.(\ref{eq5}) we have
\begin{equation}
\label{eq12}
t_{as}=\frac{1}{2\sqrt{\pi}}\frac{L}{v_0}\frac{M}{m}=\frac{1}{2\sqrt{\pi}}\frac{L}{v_0}N\mu
\end{equation}
and 
\begin{equation}
\label{eq14tth}
t_{th}=\frac{\pi}{4 N} t_{as}.
\end{equation}
Since $N>>1$, the asymptotic time $t_{as}$ is much larger 
than the thermalization time $t_{th}$, which justifies the  replacement in Eq.(\ref{eq3})
of $\dot{X}^2$ with its thermal value given by Eq.(\ref{eq10}).
It is worth noting that the difference between the time scales of
$\langle x^2 \rangle$ and $\langle \dot{x}^2 \rangle$
(i.e. $t_{as}>>t_{th}$) is associated with the presence of
the sign ``plus'' and ``minus'' in the r.h.s. of Eq. (\ref{chandraeq})$_3$ and
(\ref{chandraeq})$_4$, respectively. In fact, it is the minus sign in Eq. (\ref{chandraeq})$_4$
which is responsible for the cancellation of the slowly-decaying
asymptotic terms in
$\langle u^2 \rangle$, while these terms, present in Eq. (\ref{chandraeq})$_3$, dictate
the long time behavior of $\langle x^2 \rangle_{as}$.

The asymptotic  mean-square displacement  $\langle x^2\rangle_{as}$  of  the  piston from  its central equilibrium position $x=0$ is, according to Eqs. (\ref{chandraeq}) and (\ref{eq5}),
\begin{equation}
\label{eq16}
\langle x^2\rangle_{as} =\frac{\mu}{2}\left( \frac{L}{2}\right)^2=\frac{1}{2N}\frac{M}{m}\left(\frac{L}{2}\right)^2\text{.}
\end{equation}
In the limit $\mu<<1$ (that is,  small piston mass with respect to gas
mass), the root-mean square displacement $\langle x^2\rangle_{as}^{1/2}$ is much smaller than $L/2$,  so that both assumptions justifying  the linearization of
Eq.(\ref{eq6}), and thus the application of Langevin's approach, are  verified.  
These fluctuations can be compared with those pertaining to a
diathermal piston, 
associated with a system identical with the one described above but
for the presence 
of a thermally conducting piston instead of an adiabatic one.
To this end, let us assume the piston in Fig.1 to be a good heat
conductor, 
so that both sections possess the common constant temperature $T_0$. 
Any displacement $x$ of the piston from the central position results 
in an unbalance  $p(x)$ between the pressures $p_A(x)$ and $p_B(x)$ 
in the two sections, i.e.,
\begin{equation}
\Delta p(x)=p_A(x)-p_B(x)=\frac{N k T_0}{\mathcal{A}(L/2+x)}-\frac{N k
T_0}{\mathcal{A}(L/2-x)}\text{,}
\end{equation}
where $\mathcal{A}$ 
is the transverse section area, so that the force $F(x)$ exerted by the gas on the piston reads
\begin{equation}
\label{eq16F}
F(x)=\frac{N k T_0}{L/2 + x}-\frac{N k T_0}{L/2-x}\cong - \frac{2 N k T_0}{(L/2)^2}x\text{,}
\end{equation}
having assumed $|x|<<L/2$, as verified {\it a posteriori}. Therefore, the
piston feels both the thermal bath at temperature $T_0$ and the harmonic
potential
\begin{equation}
V(x)=\frac{N k T_0}{(L/2)^2}x^2\text{,}
\end{equation}
and, as a consequence, its position probability-density $P(x)$ reads
\begin{equation}
P(x)=\sqrt{\frac{N}{\pi (L/2)^2}}\exp\left[-\frac{ N}{(L/2)^2}x^2\right]\text{.}
\end{equation}
The associated mean square value $\langle x^2 \rangle$ is
\begin{equation}
\label{eq19}
\langle x^2 \rangle =\frac{1}{2 N}\left(\frac{L}{2}\right)^2\text{,}
\end{equation}
smaller than the corresponding mean square displacement of the
adiabatic piston 
(see Eq.(\ref{eq16})) by a factor $m/M$. 
Note that Eq.(\ref{eq19}) 
agrees with the standard result obtained in the frame 
of equilibrium-thermodynamic microscopic fluctuations. \cite{NewCallen}

Our derivation clarifies the basic difference between adiabatic and diathermal situation.
In the first case, the restoring force acting on the piston is given
by 
$-(8 N k T_0 m/L^2M)x$ (see Eq.(\ref{eq4})), 
while, in the second it reads $-(8 N k T_0/L^2)x$ 
(see Eq.(\ref{eq16F})), 
so that the restoring force acting on the adiabatic piston is $m/M$ times smaller than that acting on the diathermal one. 
We stress that the anomalously large fluctuations of the adiabatic
piston are not limited 
to the macroscopic regime, but are also present in the mesoscopic
regime, 
provided the number of molecules $N$ is large enough, as in our
case, to justify 
the use of ordinary concepts of pressure and temperature.  
This accounts for the peculiarity of the adiabatic piston system. The
large random displacements of the piston 
are not trivially related to the system dimensions, but to its
adiabatic nature: 
the diathermal piston exhibits  a mean-square-value $\langle
x^2\rangle=(m/M)\langle x^2\rangle_{as}$  
(see Eq.(\ref{eq19}) and Eq.(\ref{eq16})) smaller by a factor $m/M$ than the corresponding factor for the adiabatic piston.
\section{Variation of the entropy}
The peculiar  behavior of the adiabatic-piston  system discussed above
has a natural  counterpart in its entropy variations.  
In order to  clarify this point, we note that  the entropy behavior of
our system is not uniquely defined. First, let us consider an ensemble
of identical systems, in each  of which the piston is let indefinitely
free to move starting from the central position $x=0$ at time $t=0$. After
a  time   interval  reasonably   larger  than  $t_{as}$,   the  probability
distribution of the piston  position will stay unchanged. Accordingly,
it  is  possible to  define  an  entropy  of the  system  (gas+piston)
ensemble  whose value  turns out  to be  larger than  the  initial one
pertaining to  the one in which  all pistons are fixed  in the central
position.  The above  entropy does not allow for  a simple statistical
interpretation  like Boltzmann's one.  The latter requires  a single
system  in thermal  equilibrium,  and this  is  not the  case for  the
wandering  adiabatic  piston. Thus,  to  give  meaning to  Boltzmann's
entropy, we consider  a single system and stop the  piston either at a
given time of the order of $t_{as}$  or at a given prefixed position of the
order $\langle x^2 \rangle_{as}^{1/2}$.  
The two procedures are conceptually quite
different.  In fact, halting  the piston  at a  given time  requires a
device comparable in size  with the piston-position fluctuations it is
trying to  harness: in  other words, the  stopping mechanism  seems to
behave  like  a demon  device  while  it  rectifies the  large  piston
fluctuations, thus forbidding any violation of the Second Law. Similar
arguments  go back  to  Smoluchowsky (see  ref. \cite{Smolu}), and an interesting
example is  connected with the Feynman's pawl  and ratchet
device. \cite{FeynmanBook} 
Vice versa, in  principle, if the halting device is placed
at a position $\bar x$ chosen a priori (e.g., $\bar x=+\langle x^2 \rangle_{as}^{1/2}$), the piston stopping
operation takes place without the intervention of relevant statistical
fluctuations, i.e., without any demon mechanism. Of course, due to the
random nature of the piston wandering, we do not know when the halting
process will take place. However, if  we wait long enough (e.g., for a
time of the order  of a few $t_{as}$), there is a  high likelihood that the
piston  has  come to  a  halt at  the  prefixed  position.  The  above
argument  highlights  a  peculiar  feature  of  the  adiabatic  piston
system.  Unlike  other  systems   in  which  unusually  large  thermal
fluctuations  cannot  be   rectified  without  introducing  comparable
external  fluctuations,  the  piston  wandering  can  be  conveniently
exploited.  If the  piston  is  stopped at  a  prefixed position,  the
Boltzmann  gas  entropy significantly  decreases  without any  entropy
increase  of the  environment. To clarify this point, we 
consider  the entropy  change
$\Delta S$
undergone by  the system  when passing from  the equilibrium  state in
which the piston  is held in the central position to  the final one in
which the piston is stopped at the position $x=\bar x$.  Recalling that the
pressure does not change during the process (see Sect.II), we have
\begin{equation}
\label{DeltaXX}
\Delta S=\Delta S_A+\Delta S_B=n  c_p \ln \frac{L/2+\bar x}{L/2}+n c_p
\ln \frac{L/2-\bar x}{L/2}=
n c_p \ln \left[1-\left( \frac{\bar x}{L/2}\right)^2 \right]
\end{equation}
where  $c_p$ is  the molar  heat at  constant pressure  and $n$  the common
number of molecules in sections A and B. By recalling Eq.(\ref{eq16}) and that
$\bar x^2/(L/2)^2=\langle x^2\rangle_{as}/(L/2)^2$ is a small number, Eq. (\ref{DeltaXX}) yields
\begin{equation}
\label{eq22bis}
\frac{\Delta S}{k}=\frac{n c_p}{k} \ln \left[1-\left(\frac{\bar x}{L/2}\right)^2
\right]\cong -\frac{n c_p}{k} \left(\frac{\bar x}{L/2}\right)^2=-\frac{c_p}{2
R}N \mu\text{,}
\end{equation}
or,   by  comparing   it  with   the  standard   entropy  fluctuations
$\Delta S_{th}=(k n c_p)^{1/2}$ \cite{Landau}
\begin{equation}
\label{deltaS3}
\frac{\Delta S}{\Delta S_{th}}=-\frac{1}{2}\sqrt{\frac{c_p}{R}}\sqrt{N} \mu\text{.}
\end{equation}
This  corresponds, whenever $\mu \sqrt{N}>>1$, i.e. $M/m>>\sqrt{N}$,
to  a  sensible negative  entropy  decrease, and 
constitutes a  violation of the Second  Law, consistent with
the \textit{perpetuum mobile} behavior of  the system. 
In particular, if the system is embedded in a thermal
bath  at temperature  $T_0$,  it is  possible to
devise a  cyclic process by  means of  which the work  
\begin{equation}
\label{workeq}
W\cong k T_0 N \mu
\end{equation} 
can  be extracted  from the environment (see Appendix).
Actually, according to the accumulated experimental evidence 
confirming the validity of the Second Law, 
we do not expect these results to hold true 
in standard macroscopic situations. 
In fact, our model imposes strict limitations
on the spatial scale over which our results can be trusted. In order
to assess these limitations, we refer to the
specific case 
of a gas at standard conditions of temperature and pressure. In this case,
expressing hereafter $L$ in microns, and assuming the volume 
of the piston to be $L^3$, $N=3 \times 10^7 L^3$ it follows 
from 
Eq.(\ref{deltaS3}) that
\begin{equation}
\label{deltaS4}
\frac{ \Delta S }{\Delta S_{th}}=-5\times 10^3 L^{3/2} \mu\text{,}
\end{equation}
while from Eq.(\ref{eq12}) we have, taking $v_0=4\times10^8$ microns/sec (molecular
oxygen),
\begin{equation}
\label{deltaS5}
t_{as}=5\times10^{-2} \mu L^4 \text{sec.}
\end{equation}

We now observe that,  if we reasonably require the work $W$ extracted in
a cycle to scale with the system dimensions, i.e., with $N$, the
factor $\mu$
has to be kept constant when varying the system dimensions. Therefore,
Eq. (\ref{deltaS5}) reveals an extremely sensitive fourth-power dependence of $t_{as}$
   on the linear dimension of the system, which severely limits the
	applicability of our model to values of $L$ up to a few
       microns. 
In  fact, beyond  this mesoscopic
scale,  $t_{as}$ becomes  so  large  as to invalidate our results.  
As an  example,  for $\mu=0.01$ and  $L=1cm$, one  gets
$t_{as}=5\times10^{12} sec$,  that is about  1000 centuries! Conversely,  by taking
$L=1\mu m$, we get for $t_{as}$ a reasonable value of $5\times10^{-4}$ sec, while
$ \Delta S /\Delta S_{th}=  -50$ is  still  quite large.   
It  is important  to stress that this specific  mesoscopic regime, 
around $1\mu m$, has
emerged spontaneously  from the adiabatic-piston  problem, without the
introduction of  any \textit{ad hoc}  parameter. The Second Law,  stating that
\textit{the  entropy  of  an  isolated  system cannot  decrease},  is  actually
violated  in  the  specific  example provided  above.  The  associated
spatial scale appears to set a borderline between the macroscopic
realm and the mesoscopic one.
%%%%%%%%%%%%%%%%%%%%%%%%%%%%%%%%%%%%%
\begin{figure}
\centerline{\includegraphics[width=8.3cm]{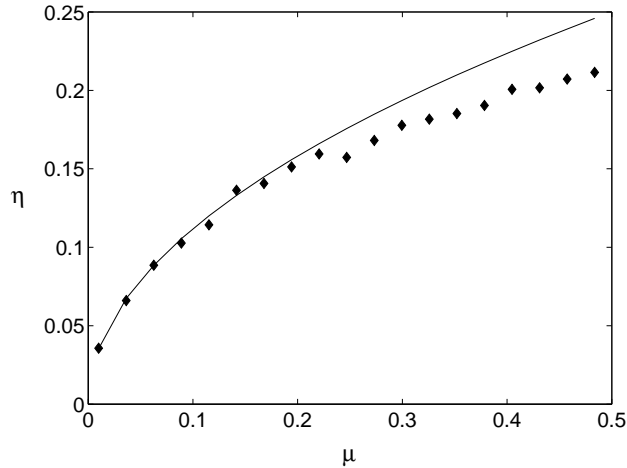}}
\caption{
Normalized asymptotic standard deviation $\eta$ of the piston position 
as a function of $\mu$ and $N=3\times10^4$
(averaged over 1000 realizations). The continuous line represents the predicted 
behavior in the linearized regime (see Eq. (\ref{eq16})).
\label{f5}} 
\end{figure}
%%%%%%%%%%%%%%%%%%%%%%%%%%%%%%%%%%%%%
\section{Beyond the linear regime}
The above approach has allowed us to deal with the
situation $\mu \ll 1$. In order to have an insight into the
behavior of our process in the more general case $\mu \lesssim 1$,
we assume Langevin's approach to be approximately valid also in
this moderately nonlinear regime, and use the
nonlinear Eq.(\ref{eq3}) with the same stochastic acceleration
$a(t)$ worked out in the linear case. After introducing the
dimensionless units $\xi
 =X/L$ and $\tau=t/t_o$, where $t_o=4 \sqrt{2} t_{th}$, Eq.(\ref{eq3}) reads
\begin{equation} \label{eq20}
\frac{d^2 \xi}{d \tau^2}+ \left( \frac{1}{\sqrt{\xi}}+\frac{1}{\sqrt{1-\xi}} \right) \frac{d\xi}{d\tau}- \frac{1}{\mu}
             \left( \frac{1}{\xi}-\frac{1}{1-\xi} \right) \left(\frac{d\xi}{d\tau}\right)^2 = \sigma \alpha (\tau),
\end{equation}
where $a(\tau)$
	 is a unitary-power white noise process and $\sigma^2
   =(\pi/2\sqrt{2})(\mu/N)$.  This last equation can be numerically
 integrated by adopting a second-order leap-frog algorithm as the one
		developed in reference \cite{Qiang00}.   

In particular, we can
   evaluate the asymptotic value of the mean-square root deviation
			    $\eta\equiv\langle (\xi-1/2)^2\rangle^{1/2}$  as
 a function of $\mu$    and, for a given value of $\mu$ , the time evolution of
		    $\langle \left(d\xi/d\tau\right)^2\rangle$.  
Fig.2  shows that the asymptotic mean-square root displacement of the
piston increases as 
$\sqrt{\mu}$
for $\mu\lesssim 0.3$  , in fairly good agreement with the
 linearized theory (see Eq.(\ref{eq16})), while it tends to saturate for larger
   values of  $\mu$. Fig. 3 confirms that the piston velocity attains its
 thermal value in a time much shorter than the asymptotic time $t_{as}$, a
necessary condition for the application of Langevin's approach. As far
as the entropy change  $\Delta S$ is concerned, the linear dependence on
$\bar x^2 =\langle x^2 \rangle_{as}$  predicted by Eq.(\ref{eq22bis}) implies as well its saturation for increasing
$\mu$ . The above results show that, for fixed N, both piston asymptotic
displacement and entropy decrease saturate with increasing piston mass
   				 $M$.
%%%%%%%%%%%%%%%%%%%%%%%%%%%%%%%%%%%%%
\begin{figure}
\centerline{\includegraphics[width=8.3cm]{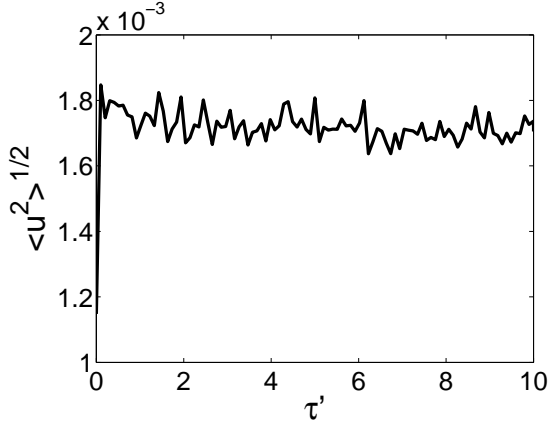}}
\caption{
Time evolution of the r.m.s. of the normalized piston velocity $u=d\xi/d\tau$ as a
function of $\tau'=10^{-4} t/t_0$, 
for $\mu=0.5$ and $N=3\times 10^4$, averaged over $1000$ realizations. 
\label{fvelocity}} 
\end{figure}
%%%%%%%%%%%%%%%%%%%%%%%%%%%%%%%%%%%%%

\section{Conclusions}
In the  present paper, we have  examined the dynamic  evolution of the
adiabatic piston system  starting  from  both   mechanical  and   thermal  equilibrium.
 Obviously, in this  case $\langle x\rangle=0$ and $\langle
 \dot x \rangle=0$ for symmetry reasons, so that the significant quantity
is $\langle  x^2(t)\rangle^{1/2}$.  This situation can  be analyzed by
generalizing  the deterministic kinetic  model developed  in \cite{Crosignani96}
with the introduction of Langevin's force.  The resulting equation for
$x(t)$ turns out  to be, whenever the mass $M$ of  the piston is small
compared  to  the  gas   mass  $M_g$,  completely  identical  to  that
describing the  Brownian motion  of an harmonically-bound  particle of
mass  $M$.  This  allows  us  to provide  an  analytic expression  for
$\langle x^2(t)\rangle^{1/2}$  , which can  be a sizeable  fraction of
the cylinder  length $L$  (see Eq. (\ref{eq16})),  thus implying  that
the piston
never reaches an equilibrium position but keeps oscillating around its
initial  position  $x=0$.  This  peculiar behavior  is  restricted  to
systems  possessing  specific  spatial  dimensions pertaining  to  a
mesoscopic realm, where the laws of thermodynamics are not necessarily
valid. The fact that the piston never stops is strictly related to the
finite  length of  the cylinder:  if we  allow the  length $L$  of the
cylinder   to  go   to   infinity  (as   assumed,   for  example,   in
\cite{Gruber03}), the ratio $M/M_g$ goes  to zero and so does $\langle
x(t)^2  \rangle^{1/2}/(L/2)$  (see  Eq.(\ref{eq16})).   At  the  same  time,
$t_{as}$ becomes extremely  large, so that, in this  limit, the piston
practically  does not  move. In  view of  this, the  violation  of the
Second Law does  not appear that startling since  it concerns a regime
outside  the  macroscopic  thermodynamic  limit. However,  it  can  be
interpreted as the signature  of something remarkable happening at the
borderline  between the macroscopic  and mesoscopic  realms, signaling
that a  new way of looking  at standard thermodynamic  concepts may be
needed if  we wish  to continue to  apply them outside  the boundaries
within which they were first  introduced.  

In order to have an insight
into the regime of  mild nonlinearity occurring when $\mu=M/M_g$ approaches
unity,  we have  numerically  solved the  stochastic  Eq.(\ref{eq20}) with  the
standard initial conditions adopted in our model. The  results of the
linearized  approach  are   in  good  agreement  with  those
associated  with the  numerical  solution of  Eq.(\ref{eq20}). 

The same problem of time evolution and approach to equilibrium of the
system  has   been  also   considered  by  using   molecular  dynamics
simulations.\cite{Kestemont00, Renne01, White02, Mansour05} These typically involve a
considerable number of point  particles (around $500$) which model the
gas inside  the cylinder and  are separated by a  frictionless movable
piston,  without  internal  degrees  of freedom,  against  which  they
undergo  perfectly elastic  collisions. Some  general  features emerge
from  the  numerical analysis,  which  reveal,  consistently with  our
results,  a very  slow approach  to  the final  equilibrium state.  

In particular, Ref \cite{White02} presents the molecular dynamics simulations
of a system evolving from an initial state in which the piston is
fixed and the number of particles and the temperatures in both
sections are equal, which is precisely the case we are dealing with in
the present paper. 
Our analytic approach predicts a relaxation time (see Eq.(\ref{eq12}))
which, in the normalized units adopted in \cite{White02} ($v_0= 2$, $m=1$ and $L=60$),
reads $t_{as}= (30/\sqrt{\pi})(M/m)\cong 17 M$. 
This is in remarkably good agreement
with the molecular dynamics simulations reported in \cite{White02} (see
Fig.5 of Ref.\cite{White02}) where, for $M<100$ (corresponding to
$\mu=M/N=100/250=0.4$), $t_{as}=16M$.

%%%%%%%%%%%%%%%%%%%%%%%%%%%%%%%%%%%%%
\begin{figure}
\centerline{\includegraphics[width=8.3cm]{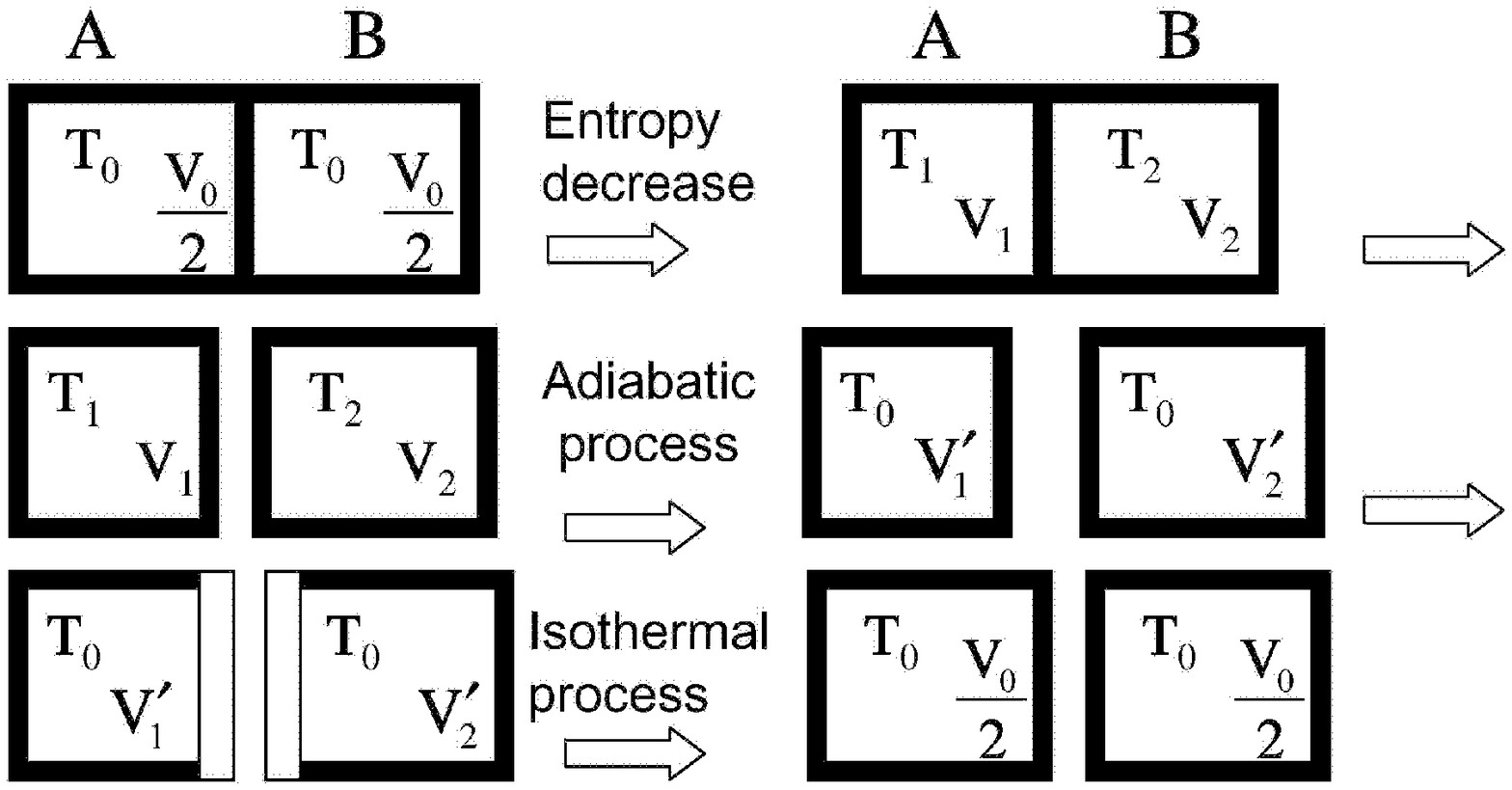}}
\caption{
\label{figcyclic}
Cyclic process through which work can be extracted from a
thermal bath at temperature $T_0$.\label{f7} } 
\end{figure}
%%%%%%%%%%%%%%%%%%%%%%%%%%%%%%%%%%%%%

In conclusion, our simple kinetic analytic description has revealed
that Callen's adiabatic piston can exhibit, when starting from a
particular mechanical and thermal equilibrium state, large
fluctuations, a very intriguing feature which presents the
characteristics of {\it perpetuum mobile}.
 This occurs for very specific
spatial  dimensions of  the system  (around  $1\mu m$),  pertaining to  the
mesoscopic regime. Whether this  is actually challenging the limits of
validity of the  Second Law seems a legitimate  question, whose answer
may shed some light on the understanding of small-scale nonequilibrium
devices.

We wish to thank Noel Corngold for his constant encouragement and many
useful comments.

%\appendix
\section*{Appendix} 
If our system is embedded in a
thermal  bath at  temperature $T_0$  and pressure  $P_0$, the following process, sketched in Fig. \ref{figcyclic}
, 
 can be
used  for extracting  work  from the  environment.  
In the  first panel, the  system is in
the initial state. The second panel 
shows the system after the entropy  decrease $\Delta S$ has occurred. 
At this point, we separate the two sections  (panel 3) and drive the two gases
back  to  the  initial  temperature  $T_0$ by  means  of  two  reversible
adiabatic  processes, the  relation between  the new  volumes $V'_1$ and
$V'_2$  and the
previous volumes $V_1$ ,$V_2$ and temperatures $T_1$,$T_2$ being
\begin{equation}
\label{eqA1}
V'_1=V_1 (T_1/T_0)^{\frac{1}{\gamma-1}}\text{,}\hspace{1cm}V'_2=V_2 (T_2/T_0)^{\frac{1}{\gamma-1}}\text{.}
\end{equation}
This is followed  (panel 4) by two reversible isothermal processes in
which  the  two sections  are  in contact  with  the  thermal bath  at
temperature  $T_0$, which  take  the  two sections  back  to the  initial
temperature $T_0$ and volume $V_0/2$  (frame 5). The total work extracted in
the process is
\begin{equation}
\label{eqA2}
W=W_1+W_2= n R T_0 \ln(\frac{V_0}{V'_1})+n R T_0 \ln (\frac{V_0}{V'_2})=
n R T_0 \ln [\frac{V_0^2 (T_0^2)^{\frac{1}{\gamma-1}}}{V_1 V_2 (T_1 T_2)^{\frac{1}{\gamma-1}}}]\text{.}
\end{equation}
Since  $V_1+V_2=2 V_0$  and  $T_1+T_2=2 T_0$,  we obviously  have  $T_0^2>T_1 T_2$  
and $V_0^2>V_1 V_2$, so that  a work $W$ larger than zero has  been obtained at the
expenses of  the heat  $Q=W$ extracted by  a single source  (the thermal
bath). It can be easily checked that $W=-T_0 \Delta S\cong k T_0 N \mu$, that
is each gas molecule contributes  with a  fraction $\mu k T_0/2$  to the
work extracted in the process.

\end{document}